\documentclass[sn-mathphys,Numbered]{sn-jnl}% Math and Physical Sciences Reference Style

\usepackage{graphicx}%
\usepackage{multirow}%
\usepackage{amsmath,amssymb,amsfonts}%
\usepackage{amsthm}%
\usepackage{mathrsfs}%
\usepackage[title]{appendix}%
\usepackage{xcolor}%
\usepackage{textcomp}%
\usepackage{manyfoot}%
\usepackage{booktabs}%
\usepackage{algorithm}%
\usepackage{algorithmicx}%
\usepackage{algpseudocode}%
\usepackage{listings}%

\theoremstyle{thmstyleone}%
\theoremstyle{thmstyletwo}%

\theoremstyle{thmstylethree}%

\begin{document}

\title[Article Title]{Helical Phononic Modes Induced by a Screw Dislocation}

\author[1,2]{\fnm{Yun} \sur{Zhou}}%\email{yuz421@eng.ucsd.edu}
\equalcont{These authors contributed equally to this work.}

\author[2,3]{\fnm{Robert} \sur{Davis}}%\email{rjdavis@eng.ucsd.edu }
\equalcont{These authors contributed equally to this work.}

\author[1,3]{\fnm{Li} \sur{Chen}}%\email{iiiauthor@gmail.com}

\author[1]{\fnm{Erda} \sur{Wen}}%\email{iiiauthor@gmail.com}

\author*[1,2,3]{\fnm{Prabhakar} \sur{Bandaru}}\email{pbandaru@ucsd.edu}

\author*[1,3]{\fnm{Daniel} \sur{Sievenpiper}}\email{dsievenpiper@ucsd.edu}

\affil*[1]{\orgdiv{Department of Electrical Engineering}, \orgname{University of California, San Diego}, \orgaddress{%\street{9500 Gilman Drive}, 
\city{La Jolla}, \postcode{92093}, \state{CA}, \country{USA}}}

\affil[2]{\orgdiv{Department of Mechanical and Aerospace Engineering}, \orgname{University of California, San Diego}, \orgaddress{%\street{9500 Gilman Drive}, 
\city{La Jolla}, \postcode{92093}, \state{CA}, \country{USA}}}

\affil[3]{\orgdiv{Program in Material Science}, \orgname{University of California, San Diego}, \orgaddress{%\street{Street}, 
\city{La Jolla}, \postcode{92093}, \state{CA}, \country{USA}}}

\abstract{
In this study, we investigate a one-dimensional (1D) unidirectional phononic waveguide embedded within a three-dimensional (3D) hexagonal close-packed phononic crystal, achieved by the introduction of a screw dislocation. This approach does not rely on the non-trivial topological characteristics of the 3D crystal. We discover that this dislocation induces a pair of helical modes, characterized by their orthogonal $x$- and $y$-directional displacements being out of phase by 90 degrees, which results in a distinctive rotational motion. These helical modes demonstrate directional propagation, tightly linked to the helicity of the screw dislocation. Through considerations of symmetry, we reveal that the emergence of these helical modes is governed by the symmetry of the screw dislocation itself. Our findings not only provide insights into the interplay between dislocation-induced symmetry and wave propagation in phononic systems but also open new avenues for designing directionally selective waveguides without relying on the crystal's topological properties.}


\maketitle

\section{Introduction}\label{sec1}

Defects \cite{nowick_golden_1996}, inherent imperfections or irregularities in the crystal structure of materials, are prevalent in various forms. These anomalies exert a measurable impact on the physical, electrical, and mechanical properties of materials. The development of phase classification systems based on symmetry, concurrent with the exploration of topological properties of matter \cite{thouless_quantized_1982,qi2011topological}, has spurred a subsequent deluge of findings that establish links between crystalline defects and topological phenomena.

Various types of defects are termed "topological" as they represent local disruptions in the underlying lattice structure that cannot be eliminated through any smooth transformation \cite{mermin_topological_1979}. The screw dislocation, which is the focus of the current investigation, exemplifies this characteristic. Away from the screw axis, the lattice appears unperturbed, and the local distortion can not be rectified without altering lattice sites at arbitrarily long distances from the dislocation line. Meanwhile, the broad class of topological defects also include those that give rise to robust states existing on the defects themselves \cite{lin_topological_2023}. These states rely upon various formulations of the bulk-boundary correspondence \cite{hatsugai_chern_1993}, where a nontrivial value of a topological invariant (such as the Chern number or $\mathcal{Z}_2$ index) implies the existence of trapped interface modes within the defect. 

Defects of such form have recently been demonstrated to have topological features in many material systems \cite{lin_topological_2023}, which enforce the existence of modes that are robust to scattering phenomena. %that do not close the bulk bandgap.
Disclinations in 2D, where a sector of the bulk crystal over an angle (called the Frank angle $\Omega$) is removed and the remaining bulk "glued" back together by connecting the two newly created edges (the Volterra process \cite{kleman_disclinations_2008}), have shown a rich array of experimentally demonstrable signatures of topologically robust states \cite{wang2020observation,liu2021bulk,peterson2021trapped}. Edge dislocations in 2D, a line defect where the arrangement of atoms abruptly terminates, have been shown to form 0D topological cavities in electromagnetic systems \cite{li2018topological}. The range of defects within 3D systems is considerably larger than these 2D settings, but present difficulties both in theoretically describing their behavior as well as in building a suitable experimental platform. 

One example of topological defects in 3D systems is a one-dimensional (1D) topological waveguide in a 3D structure trapped at a defect \cite{ran2009one}. It was theoretically shown that by introducing a screw dislocation into a 3D topological material, robust 1D topological edge states appear at the center of the dislocation. Such states are protected by a weak topological index, which characterizes the underlying reciprocal-space topological behavior of the bulk, and a Burgers vector, which uniquely defines the screw. Similarly, partial dislocations at stacking faults, where the defect shows signatures of the lattice deformation at all length scales, have also been shown to host topologically robust states \cite{queiroz2019partial}. 

While these lattice defects are common in natural materials, probing these topological properties is generally restricted to engineered materials in electromagnetics or acoustics. A screw dislocation that gives rise to a 1D topologically robust mode in a photonic/acoustic system can be formed in a weak 3D topological insulator composed of a stack of decoupled 2D topological insulators\cite{ye2022topological,xue2021observation,lustig2022photonic,lin2018three}.
Other ideas to implement 1D topological waveguides in 3D systems include constructing a defect line at the center of a disclination\cite{wang2021vortex} and building second-order hinge states\cite{kim2020recent,wei2021higher,wei20213d,he2023excitation,pu2023acoustic} in 3D Weyl crystals. However, the difficulty of creating viable experimental platforms for these engineered materials has made realizing 1D topological modes challenging.
 
In all previous studies discussed above, the initial starting point is always a 2D or 3D topologically nontrivial  material, and 1D edge states are induced within the bandgap via an interplay of the defect and the preexisting topology \cite{lin_topological_2023}. An open question has been can we find a robust 1D state in a \textit{trivial} 3D crystal? 

Here, inspired by topological waveguides created by defect lines in 2D seemingly trivial systems  \cite{zhou2023chip,kong2020spin,davis2022topologically}, we propose and experimentally demonstrate a 1D topological waveguide in a trivial 3D hexagonal close packed (HPC) phononic crystal by introducing a screw dislocation in the lattice. We show via simulations and direct measurement that a pair of helical modes exist at the position of the screw dislocation, which proves that unidirectional edge states can exist via structural helicity, without the requirement of a 3D topologically non-trivial system. The structural behavior is simple to realize and probe via an additively manufactured lattice. Our findings can be potentially used in applications such as vibration damping \cite{chung_review_2001} and unidirectional sound propagation for privacy protection \cite{sagartzazu_review_2008}. 

\section{Results}\label{sec2}
To show how the screw dislocation results in robust modes, we first analyze the underlying HCP structure computationally in sections \ref{subsec1} and \ref{subsec2}, then demonstrate the acoustic behavior experimentally in section \ref{subsec3}. 

\subsection{Helical Phononic Modes in a HCP Lattice}\label{subsec1}
\begin{figure}[h]
\centering
\includegraphics[width=0.9\textwidth]{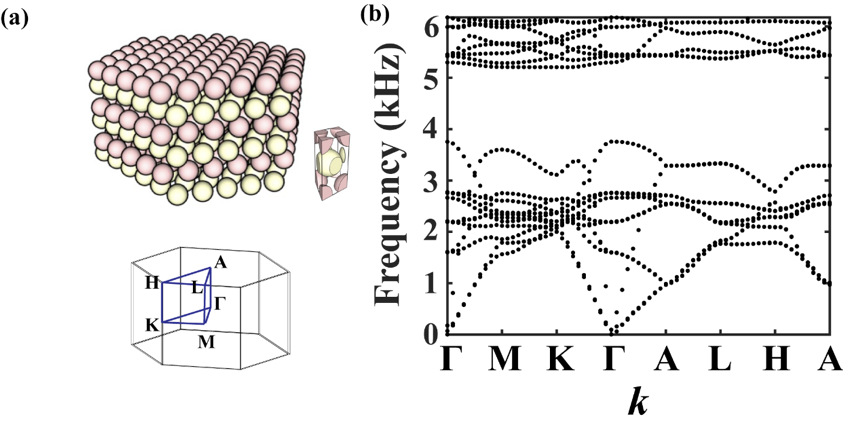}
\caption{HCP phononic crystal with a band gap. (a) HCP phononic crystal made with stainless steel balls in a soft rubber background material. The metal balls have a density of 7780 $\text{kg}/\text{m}^3$, Young's modulus of 207 GPa, and Poission's ratio of 0.279. The background material has a density of 1180 $\text{kg}/\text{m}^3$, Young's modulus of 3.9 MPa, and Poission's ratio of 0.48. The two colors indicate the AB stacking of the HCP crystal. (b) Band dispersion with a 3D band gap in the HCP crystal.}\label{fig1}
\end{figure}

We begin with a 3D phononic crystal with a band gap – a HCP crystal made with solid spheres embedded in a elastic background material, as shown in Fig.\ref{fig1}(a). Previous research has shown that a 3D phononic band gap can be formed in such a structure\cite{sainidou2002formation}, evidenced by the band dispersion curves in Fig.\ref{fig1}(b). Here, the spheres are solid carbon steel balls with diameter of $d=0.635$ cm. The HCP crystal is with in-plane lattice constant $a=1.316d$ and out-of-plane lattice constant $c=\sqrt\frac{8}{3}a$. A 3D band gap is formed between 3.75 kHz to 5.20 kHz. 

The HCP crystal structure (space group 194) is time reversal symmetric, and so lies in the AI Altland-Zirnbaur symmetry class \cite{altland1997nonstandard}. From this classification, we determine that the bandgap observed in Fig. 1(b) is "trivial" with respect to a wide range of topological analytical tools (see Sec. \ref{sec:disc}). We will show, however, that robust edge states nevertheless appear along the screw axis. 

\begin{figure}[h]
\centering
\includegraphics[width=0.9\textwidth]{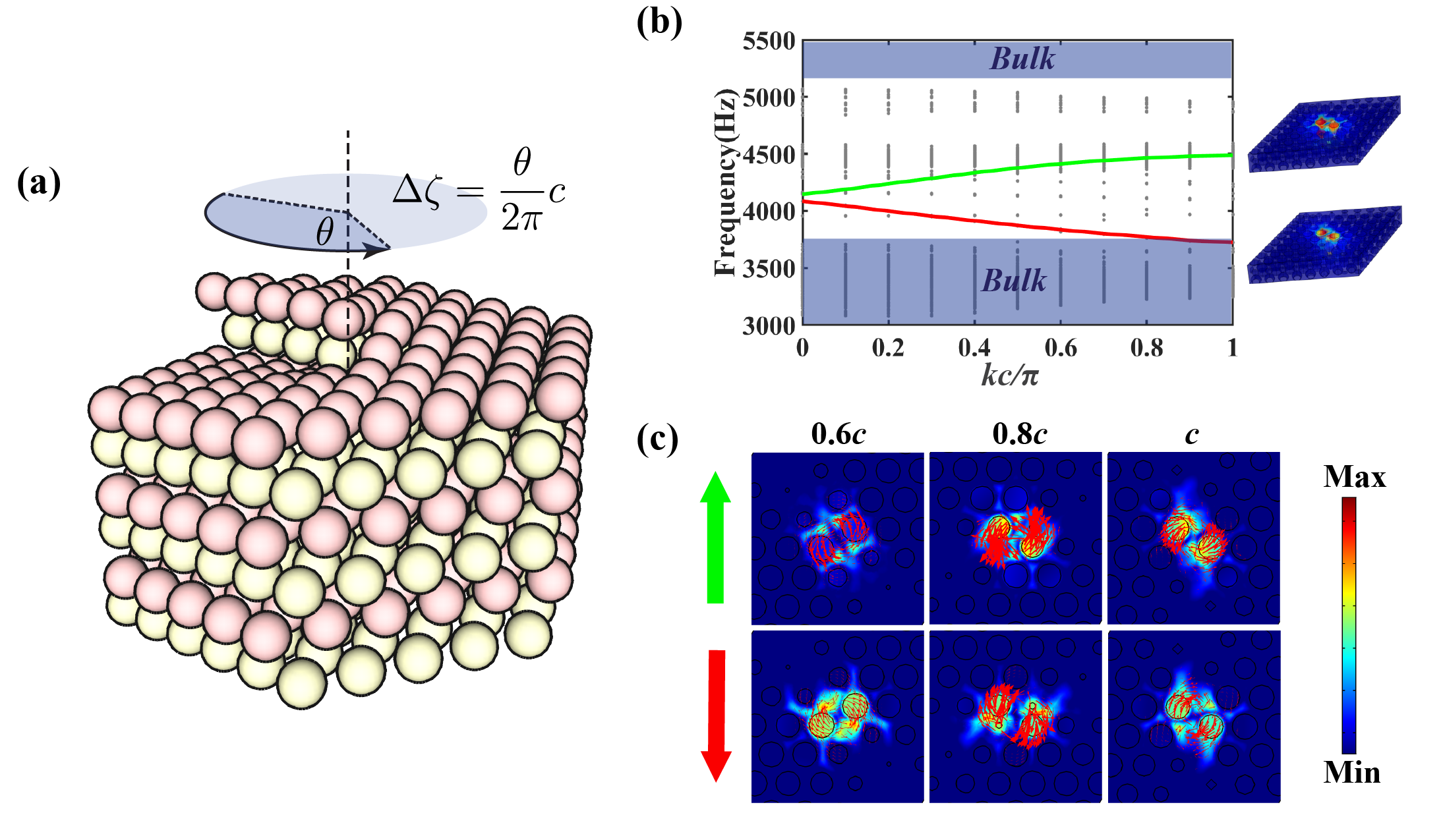}
\caption{Helical phononic modes induced by the screw dislocation. (a) HCP phononic crystal made with a screw dislocation at its center. (b) Counter-propagating helical modes in the phononic band gap at cross at $\Gamma$ (c) Cross-sections of the eigen displacement fields correspond to the counter-propagating helical modes at the heights of $0.6c$, $0.8c$ and $c$. Note that there are modes visible within the bandgap region that do not correspond to the unidirectional modes at the screw center; they are instead modes existing at the corners and edges due to the finite simulated size.}\label{fig2}
\end{figure}

We introduce a screw dislocation to the center of a supercell HCP structure with 8 unit cells in the $x$ and $y$ directions, and one in the $z$ direction ($8\times 8 \times 1)$, with Burgers vector $\mathbf{b}$ perpendicular to the $xy-$ plane with the length equal to the out-of-plane lattice constant $c$, as shown by Fig. \ref{fig2}(a). Each metal ball in the original HCP lattice is moved by $\Delta\zeta =\frac{\theta}{2\pi}c$ in the z direction to form the screw dislocation, where $\theta$ is each ball's angle with respect to the center. The resulting supercell spectrum along the direction of the screw dislocation is shown in Fig.2(b). A pair of dispersion curves with opposite group velocities are observed in the frequency range of the bulk band gap of the HCP crystal, which cross at the $\Gamma$ point. The cross sections of their eigen-displacement fields at $0.6c$, $0.8c$ and $c$ are shown in Fig. 2(c), with arrows indicating the direction of the in-plane energy flow. It can be observed that these modes are confined at the center of the introduced screw dislocation and their energy flows in a rotational manner with opposite chirality, which suggests the directionality of these modes.

We note that there are also modes in the HCP band gap that correspond to boundary or corner states that appear within Fig. \ref{fig2}(b), which are unrelated to the central screw axis modes. They are caused by the finite size of the supercell simulated in the eigen-mode simulation and are not due to the screw dislocation. 

\subsection{Directionality of the Helical Modes}\label{subsec2}
\begin{figure}[h]
\centering
\includegraphics[width=0.9\textwidth]{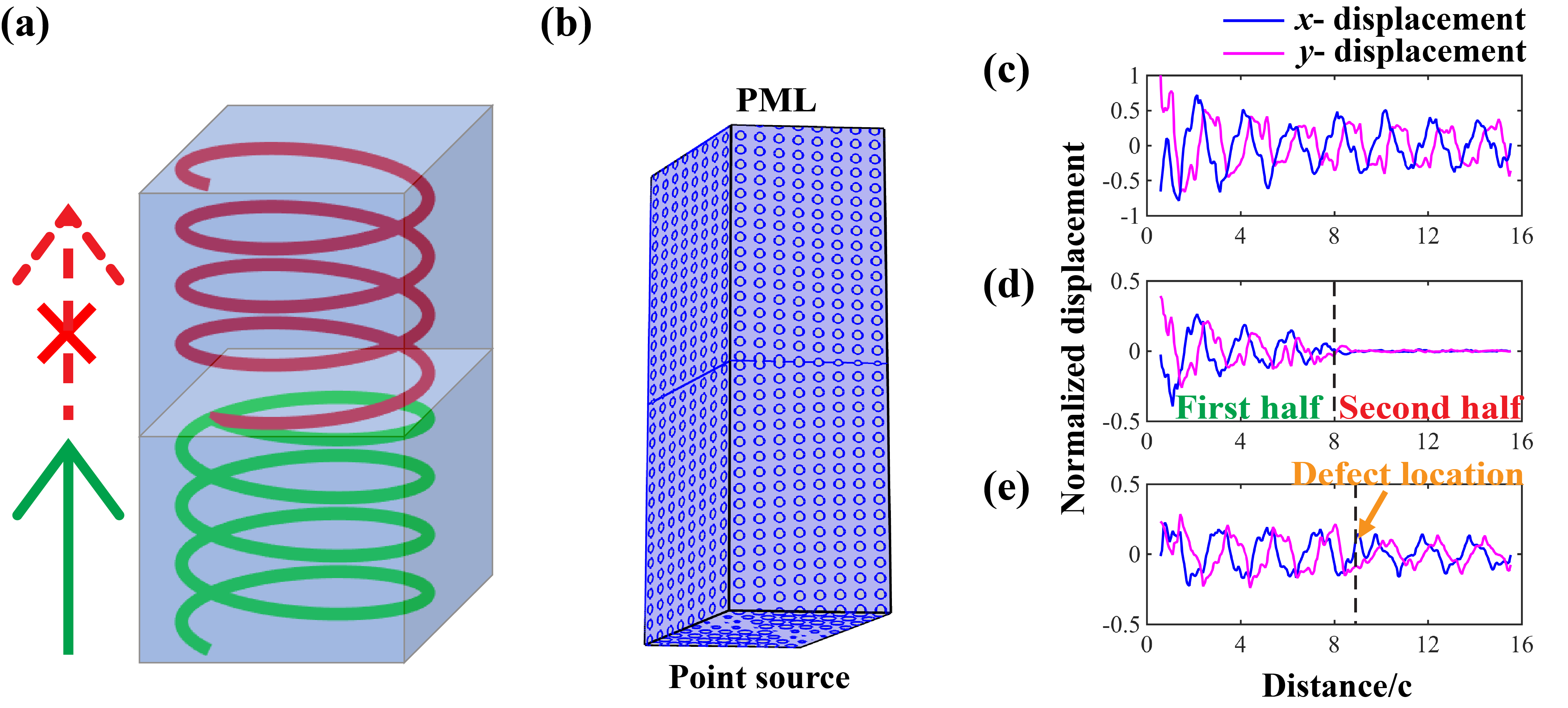}
\caption{Unidirectional helical phononic modes. (a) Schematic view of two waveguides with two screw dislocations with opposite helicity connected. (b) View of the 3D HCP lattice that realizes the structure in (a). A the bottom the acoustic point source is placed to excite the mode, while the top surface is a perfectly matched later (PML) to prevent backwards scattering. (c) Spatial displacement along the waveguide with one helicity only, (d) opposite helicities, and (e) one helicity but with a defect along the waveguide. The base case in (c) shows the displacement being near constant, while the case of flipped helicity in (d) shows near zero transmission past the interface at the middle. (e) Illustrates the robustness to minor scattering.}\label{fig3}
\end{figure}

The existence of the in-gape modes seen in the eigenvalue spectrum of Fig \ref{fig2}(b) suggest unidirectional edge states reminiscent of other topological structures \cite{lin_topological_2023}, but cannot be clearly distinguished from ordinary defect modes without further investigation. To demonstrate the directionality of these modes, we construct a waveguide composed of two screw dislocations with opposite helicity connected to each other, as shown schematically in Fig. \ref{fig3}(a), and compare with a waveguide of the same length that is composed of one screw dislocation. If the bandgap-localized modes were merely defect modes, we would expect the transmission to be similar between these two cases, with the only difference being finite scattering occurring at the interfacial discontinuity. 

In the simulation, we excite the waveguides at their bottom side, with a point source of frequency $f_e=4.414$ kHz within the 3D bandgap where the helical modes are found, and place a perfect matching layer on the top of the waveguide to eliminate any reflection\cite{berenger1994perfectly}, as shown in Fig. \ref{fig3}(b). 

The in-plane displacement along the center line of the single helicity waveguide is shown in Fig. \ref{fig3}(c). We find that the mode propagates with minimal attenuation, and that the $x$- and the $y$- displacements along the waveguide always have a 90-degree phase difference, a clear indication that the spatial displacement at the dislocation is rotating along the waveguide.

In contrast, the in-plane displacement along the center line of the stack of two different helicity waveguides, shown in Fig. \ref{fig3}(d), shows a significant loss of amplitude at the interface between the two helical structures. This suggests the coupling between the forward and backward helical modes is low, on the order of -20 dB for the case shown. Behavior of this type is similar to that seen in robust edge states in topological waveguides, which have ideally zero coupling of the two directions of propagation, even in time-reversal symmetric systems \cite{bisharat_photonic_2021}. We can contrast this with the behavior between two pure defect modes connected together, whose coupling is determined by the scattering at the interface and result in large forward transmission \cite{waks_coupled_2005}. 

 To demonstrate the robustness of the waveguide, a metal ball along the center of the waveguide with one helicity is removed, and its in-plane displacements are plotted in Fig. \ref{fig3}(e). Robust wave propagation is still observed with the defect introduced, and the wave continues to propagate with a constant magnitude after it passes the defect.

\subsection{Measurement results}\label{subsec3}
\begin{figure}[h]
\centering
\includegraphics[width=0.9\textwidth]{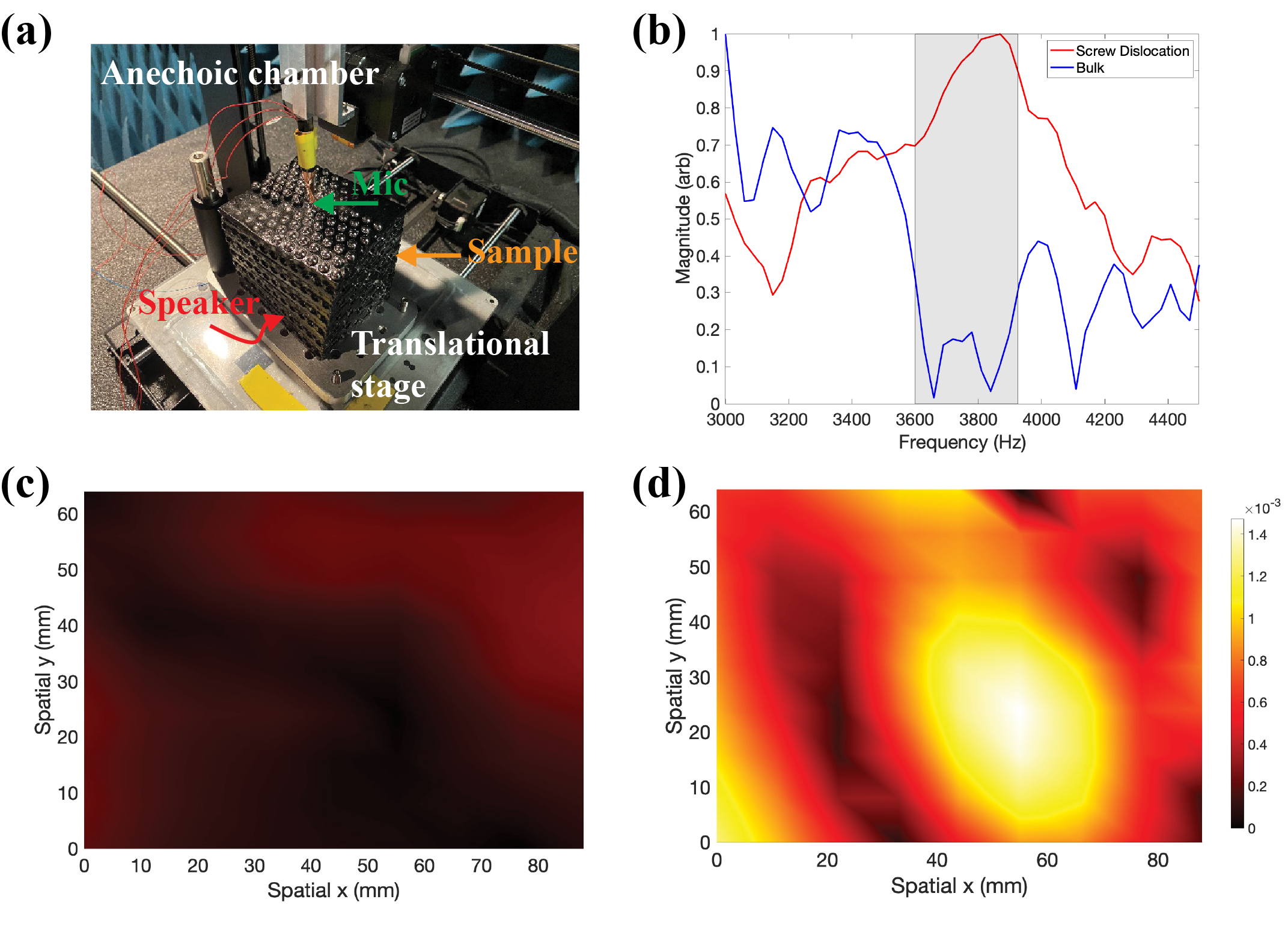}
\caption{Measurement of the waveguide containing a screw dislocation. (a) Measurement setup. The sample sits on a 2D translation stage, with a fixed speaker source below and the microphone receiver moved across the top surface. (b) Normalized magnitude of the microphone signal at the center of the waveguide for the bulk and screw dislocated samples. (c) and (d) are the field scan on the top of the bulk and screw dislocated samples at frequencies in the bandgap, respectively. The color scaling for each is equal, which shows that the bulk sample in (c) has negligible transmission over the whole surface, while there is a strongly peaked mode trapped near the center visible in the screw dislocated sample (d).}\label{fig4}
\end{figure}
The simulation results in the previous section indicate robust, unidirectional edge states propagating within the bandgap of the HCP lattice. To verify this experimentally, we prepared a sample of the HCP crystal with the proposed screw dislocation at its center, as well as a reference HCP crystal without the screw dislocation, with both made of identical materials as those used in the simulations of the previous section. The background material (Tango Black simulated rubber) has low acoustic impedance compared with the scattered spheres ($Z_{background}\approx1/20 * Z_{spheres}$), and is 3D printed layer-by-layer. Solid 440 stainless steel metal spheres are press fitted into the 3D printed background matrix to build the samples.

We obtain the intensity of the sound field by measuring the pressure field displacement amplitude. The measurement setup is shown in Fig. \ref{fig4}(a). The measurements are performed using an automated translation stage. We excite the acoustic wave by a directional speaker, which is placed at the center of the bottom of the sample. The receiver above the sample is a microphone that is inserted into a metal tube with diameter of 2 cm to improve spatial resolution.  A lock-in amplifier is used to capture the signal measured from the microphone, with an internally generated sine wave used as the input to the speaker, to synchronize the frequencies from the source to the receiver. A total of 81 points are measured across a grid of 9$\times$9 locations, with spatial resolution of 11 mm in the $x$- direction and 8 mm in the $y$- direction. 16 measurements are averaged for each data point. All measurements were performed inside an anechoic chamber to avoid interference from other noise sources. The 2D stage moves the microphone 0.5 cm above the top surface of the sample during scanning. 

A transmission frequency sweep over the expected bandgap region was performed for both samples with (red) and without (b) defects, the results of which are shown in Fig. \ref{fig4}(b). The range of the bandgap is indicated on the figure. For both samples, the receiver was placed in the geometric centers, which aligns with the screw axis for the nontrivial sample. All measurement data are normalized by the measured free-space transmission (with the microphone directly connected to the speaker), and each set of data is then normalized to their respective maxima. We observe a clear mode existing within the expected bandgap region, with the sample with the screw dislocation showing high transmission at the lowest transmission region of the bulk sample's bandgap. This behavior closely follows the expected behavior from Fig. \ref{fig2}(b). 

To verify the spatial localization of the mode, a 2D scan over the top surface of both samples was performed, with the frequency taken at the highest point of transmission for the sample with a screw dislocation in Fig. \ref{fig4}(b). From Fig. \ref{fig4}(d) we see that sample with the defect shows a mode confined at the screw dislocation near 3.8 kHz, whereas the bulk sample in Fig. \ref{fig4}(c) shows uniformly low transmission, as expected. Note that both 2D surface plots are set to equal, non-normalized units. 

\section{Discussion}\label{sec:disc}
The above discoveries from the simulations and measurements raise a question: if the system is topologically trivial, how do the helical modes appear in the screw dislocation? Past studies have shown that screw dislocations do indeed result in nontrivial modes\cite{ran2009one}, but these efforts have focused mostly on systems with spin degrees of freedom in the AII Altland-Zirnbaur class \cite{altland1997nonstandard}, and critically rely on the existence of spin-orbit coupling \cite{hu2018ubiquitous}. Likewise, many past studies have shown that stacks of nontrivial 2D topological insulator models, even in class AI (e.g., spin-1 systems with time reversal symmetry), may possess nontrivial modes at dislocations \cite{wang2021vortex,lustig2022photonic,ye2022topological}, but still require the 2D system’s nontriviality. Such platforms work by a weak coupling of the nontrivial 2D sheets, where robust edge states may propagate along the dislocation. In such systems the screw dislocation acts as a 1D edge that cuts through the bulk, connecting the symmetry-protected edge states on each sheet \cite{ran2009one}.

In the present HCP lattice, there is both time reversal symmetry and inversion symmetry, which precludes any Berry phase-related phenomena \cite{liu2017novel}. Hence, we may conclude the helical states are not the result of coupled 2D sheets of topological edge states. 

Another possibility is that of crystallographic symmetries \cite{fu2011topological}. Recent developments in the field of topological systems include the powerful tools of symmetry indicators \cite{po2017symmetry} and topological quantum chemistry \cite{bradlyn2017topological}. These methods exploit the combinatorial and symmetry restrictions placed upon physical systems to characterize what nontrivial modes can possibly exist. However, we rule this out, given knowledge that the symmetry indicator group for spinless systems with TRS in space group 194 (that of the HCP used) is given by $\mathbb{Z}_1$, which is trivial \cite{po2017symmetry}. 

Since the symmetry indicator (as well as all other topological markers) imply triviality of the bulk crystal, we must examine further how the introduction of a screw dislocation influences the system. A screw dislocation is a topological feature in real space, being irremovable for adiabatic changes unless a separate screw dislocation with opposite helicity is combined with it. Moreover, at sufficient distances from the screw axis the crystal becomes indistinguishable from the bulk material. 

In Ref. \cite{ran2009one}, an index was derived that links the Burgers vector $\mathbf{b}$ with the so-called weak index $G_v$ (which is analogous to a Chern number for weakly coupled 2D sheets along a given direction) to determine the existence of nontrivial modes along a screw axis, given as $\Theta = \frac{1}{2\pi}\mathbf{G_v} \cdot \mathbf{b}$. When this value is nonzero, a pair of helical modes will be trapped in a bulk bandgap. However, this index assumes a nonzero value of the weak index, which, in the case of systems in AI with time reversal symmetry, are zero. 

We may instead observe that the introduction of the screw dislocation into the otherwise topologically trivial crystal structure acts to lower the allowed symmetries of any mode near the defect line. If we consider the effective spinless Hamiltonian of the bulk crystal with a screw dislocation, we may assume that, within a bandgap region of the non-defected crystal, no bulk states may appear anywhere except possibly at the 1D defect line. This is to be expected even for simple defects which do not possess any special topological properties. To see this, we consider a toy model.

We may express the full Hamiltonian of a system containing a screw dislocation by
\begin{equation}\label{eq1}
    H_{3D}(x,y,z) = \sum_{k_z} e^{ik_z z} H_{2D}^{eff}(x,y,k_z),
\end{equation}
where the phase factor $e^{ik_z z}$ implies that the crystal is periodic along the Burgers vector-aligned direction $z$, and $H_{2D}^{eff}(x,y,k_z)$ is the \textit{effective} Hamiltonian corresponding to the 2D sheet perpendicular to the screw axis \cite{slager_interplay_2014}. Note that far from the screw axis, Eq. \eqref{eq1} tends to the bulk crystal Hamiltonian (as represented for the present study by the 3D band structure in Fig. \ref{fig1}(b)) as $1/r^2$ for a distance $r=\sqrt{x^2 + y^2}$. Hence, we can guarantee that any mode within the bulk crystal bandgap must exist only within the region of the screw dislocation itself, and will decay exponentially in the perpendicular directions away from it. 

In the vicinity of the screw axis, we then define the 1D tight binding Hamiltonian along $z$ as 
\begin{equation}
    H = \sum_{\langle i,j\rangle} t c^\dagger_{i}c_{j},
\end{equation}
where the sum runs over the nearest neighbor sites $i,j$ along the screw axis, which have hopping amplitude $t$. For a fermionic system, the above TB model could then be derived by exploiting the structural symmetry imposed by the screw dislocation to arrive at an effective model via group theoretic considerations \cite{hu2018ubiquitous}. However, the vectorial nature of the phononic modes studied here complicate this process, since the three polarization components induce more degrees of freedom than the scalar Hamiltonian equation provides. 

Instead, we consider the two possible helical directions, clockwise and counterclockwise, as a basis of states along the screw axis. These two states can be clearly distinguished by the rotational circulation of the spatial displacement field components seen in Fig. \ref{fig2}(c), and act like pseudospin states $\psi^{\uparrow/\downarrow}$. $\psi^{\uparrow}$ here implies a state of one helicity (e.g., right handed) while $\psi^{\downarrow}$ implies the opposite helicity (e..g, left handed). We have from symmetry considerations that 
\begin{align}
    \psi^\uparrow (-k_z) = \psi^\downarrow(k_z),\\
    \psi^\downarrow (-k_z) = \psi^\uparrow(k_z).
\end{align}
In other words, time reversal symmetry acts to flip the pseudospin, as the state must maintain the structural symmetry of the screw dislocation. This still maintains the global bosonic time reversal symmetry $\mathcal{T}^2 = +1$, but shows the distinction between the screw dislocation modes in Fig. \ref{fig2}(b) and ordinary defect modes. This also demonstrates the lack of coupling between the two opposite helicity structures in Fig. \ref{fig3}(d), as $\langle \psi^\uparrow(+k_z) | \psi_\downarrow(+k_z)\rangle = 0$ by definition. We then conclude that the structural symmetry enforces helically-distinguishable states. 

\section{Methods}\label{sec11}
The eigenmode and the driven mode simulations were implemented using COMSOL Multiphysics with the Acoustic (Acoustic-Solid Interaction) module, based on the finite element method. Floquet periodic boundaries were assigned for unit cell and supercell band diagram calculations of Fig. \ref{fig1} and \ref{fig2}, while a perfect matching layer was imposed above to absorb reflections for the frequency domain driven-mode simulations of Fig. \ref{fig3}. In the driven mode simulation of Fig. \ref{fig3}, we excite the wave by applying a sinusoidal point source on one side of the waveguide. 

Measurements were conducted in an anechoic chamber. The background material is printed with an flexible material Tango Black by a 3D printer Stratasys Connex3. A speaker (Visaton GmbH $\&$ Co. KG, K16-50OHM, 16 mm in diameter, 3.5 mm in height) was driven by a lock-in amplifier (Stanford Research 830) to excite the sound wave. A microphone (Knowles FG-23629-P16, 2.57 mm in diameter and height) is inserted into a metal tube with diameter of 2 cm to reduce the measurement noise via sampling over a larger area. The speaker was fixed at a position 0.5 cm above the sample facing the top surface of the sample. During the measurement, the sample was moved by a 2D translation stage across its surface.

\section{Conclusion}\label{sec12}
In summary, we demonstrate a unidirectional phononic waveguide by introducing a screw dislocation in an otherwise topologically trivial HCP 3D phononic crystal made of solid metal spherical scatters in a elastic background material. We prove the existence of the helical modes by band dispersion calculations in FEM simulations, as well as transmission spectrum data from the measurements. We find that the $x$- and $y$- directional displacements have a 90 degree phase difference, which shows the rotational behavior of the modes.

Using considerations of symmetry, we show that the screw dislocated modes are not due to the standard set of topological classifications, but instead arise from the structural behavior of the screw dislocation itself. This is in sharp contrast to previous efforts, where there is always an assumption of either weakly coupled 2D systems possessing inherent topology, or the requirement of a spin-orbit coupling, neither of which is the case here.

\bmhead{Acknowledgments}
This work was supported by Army Research Oﬃce Contract No. W911NF-17-1-0453 and AFOSR Grant No. FA9550-16-10093.

\bibliography{sn-bibliography}% common bib file

\end{document}